ORIGINAL RESEARCH

**Title:** How to apply multiple imputation in propensity score matching with partially observed confounders: a simulation study and practical recommendations

**Authors**: Albee Y. Ling[1], Maria E. Montez-Rath[2], Maya B. Mathur[1], Kris Kapphahn[1], Manisha Desai[1]

**Author Affiliations:**

[1]Quantitative Sciences Unit, Division of Biomedical Informatics Research, Department of Medicine, Stanford University Medical Center, 1070 Arastradero Road, Palo Alto, CA 94304

[2]Division of Nephrology, Department of Medicine, Stanford University Medical Center, 1070 Arastradero Road, Palo Alto, CA 94304

**Abstract**

Propensity score matching (PSM) has been widely used to mitigate confounding in observational studies, although complications arise when the covariates used to estimate the PS are only partially observed. Multiple imputation (MI) is a potential solution for handling missing covariates in the estimation of the PS. Unfortunately, it is not clear how to best apply MI strategies in the context of PSM. We conducted a simulation study to compare the performances of popular non-MI missing data methods and various MI-based strategies under different missing data mechanisms (MDMs). We found that commonly applied missing data methods resulted in biased and inefficient estimates, and we observed large variation in performance across MI-based strategies. Based on our findings, we recommend 1) deriving the PS after applying MI (referred to as *MI-derPassive*); 2) conducting PSM within each imputed data set followed by averaging the treatment effects to arrive at one summarized finding (*INT-within*) for mild MDMs and averaging the PSs across multiply imputed datasets before obtaining one treatment effect using PSM (*INT-across*) for more complex MDMs; 3) a bootstrapped-based variance to account for uncertainty of PS estimation, matching, and imputation; and 4) inclusion of key auxiliary variables in the imputation model.

## 1. Background

Randomized clinical trials serve as the gold standard for providing strong evidence for the effects of new and existing treatments.[1] For numerous reasons including ethical and financial costs, however, it is not always feasible to conduct such trials. Alternatively, observational studies have a long history of providing evidence for comparative effectiveness of treatments and interventions, and can also serve as justification for conducting a definitive randomized clinical trial.[2–4] The presence of confounding, however, can threaten the ability of an observational study to draw causal inference.[5] Methods based on the propensity score (PS), defined as the conditional probability of being assigned a particular treatment given the subject's observed baseline covariates, can be used to mitigate issues of confounding present in observational studies.[6–8] While the true PS is not typically known, it can be estimated using logistic regression or other machine learning techniques.[9] PS-based methods include matching, inverse probability of treatment weighting (IPTW), stratification or subclassification, and covariate adjustment.[7,9–11] PS matching (PSM) is one of the more common tools used among the PS-based techniques and, thus the primary focus of the study presented here.

By way of background, we can estimate the average treatment effect in the treated (ATT) using PSM under the potential outcomes framework.[12,13] PSM produces unbiased estimates of the ATT, under the assumptions of strongly ignorable treatment assignment (SITA)[6], which requires 1) the exposure to be independent of potential outcomes given a set of covariates (unconfoundedness) 2) the probability of receiving each treatment conditional on any set of covariates to be strictly between zero and one (positivity), as well as the Stable Unit Treatment

Value Assumption (SUTVA)[14], which states that the outcome of a subject is not affected by the treatment assignment of other subjects. Once these assumptions are met, researchers can choose from a variety of matching methods. This paper focus on nearest neighbor matching. Other choices in matching include greedy or optimal matching, with or without replacement, one-to-one or many-to-one matching.[15] Individuals with comparable PSs and discordant exposures will be matched to achieve balance in covariates in the PS model across the comparator groups of interest. One way to ensure quality of matches is to introduce a caliper, although this introduces potential bias and inefficiency by discarding treated units outside the area of common support.[15–18]

Once balance of covariates has been achieved in the matched samples, an analysis can be conducted to estimate the treatment effect and its variance. In contrast to a simple comparison between the treatment groups within the matched samples, a regression-based treatment effect estimator removes residual imbalance in covariates between treatment groups by adjusting for confounders in the model after matching.[15,19,20] The variance estimation of the treatment effect in the context of PSM is not straightforward and remains controversial despite the large body of literature devoting attention to this issue.[15,19,21–30] In addition to the uncertainty in the treatment effect estimation, researchers disagree on how to account for uncertainty in the PS estimation[15,25] or in the matching process[21,26,29], if at all. Based on the current literature, we considered two variance estimators as relevant choices: a robust cluster variance estimator[26] to account for the clustering induced by matched observations as well as a bootstrapped-based estimator[28,31] as it takes into account uncertainties in both the PS estimation and the matching process.

The statistical validity of PSM is threatened in the presence of missing data.[11,32–34] For example, if systematic missingness exists among measured confounders, the estimated ATT may be biased. The most common approaches to handling partially observed confounders in PSM include complete-case analyses (CC), complete-variable analysis (CVA), and single imputation methods.[37] In the former, subjects missing at least one confounders are excluded from the analysis.[35,36] Importantly, CC produces unbiased estimates when data are missing completely at random (MCAR), i.e. missingness is not related to either observed or unobserved data. Complete-variable analysis (CVA) involves excluding variables with missingness from the analysis, and single imputation methods have been applied in this context although less frequently than CC and CVA.[37] Multiple imputation (MI) is a reasonably flexible method for handling missing data with good statistical properties that leads to unbiased and efficient estimators of parameters of interest when the data are missing at random (MAR), or when the missingness is related to observed data and not unobserved data conditional on the observed[38]. MI may also be applicable when data are missing not at random (MNAR), or when missingness is related to unobserved variables, although researchers either need to explicitly model the missing data mechanisms under MNAR.[39] The implementation of MI even in the simplest of contexts, however, requires that the user make numerous decisions which can greatly impact the results.[40] Among the two modelling approaches of MI, our study focuses on fully conditional specification (FCS) instead of joint modeling (JM) for its flexibility to accommodate multiple data types and its increase in application.[41]

In the context of PSM, MI presents unique issues. To incorporate the PS when using MI, one has to (1) estimate the PS and (2) integrate the PS into the analysis to obtain the treatment effect. There are multiple options for the estimation step. Specifically, it is not clear whether one should impute the confounders first and then estimate the PS, referred to as a passive approach[40], or whether one should impute the PS as if it were any other variable, referred to as an active approach[42]. The question of imputing in the presence of derived variables is not new and has been discussed in previous contexts, including for imputing interaction terms and higher-order terms;[42–46] however, the approach utilized in the context of PSM has been limited.[47–49] Active approaches have been promoted as bias-reducing because all variables and their interrelationships are considered in the imputation process, reflecting a proper and congenial imputation approach.[40,50–52] In contrast, passive approaches have been supported because they result in internally consistent imputations (where the PS for subjects will perfectly correspond to its estimation as a function of their underlying confounders). Regarding the integration of PS, one can apply PSM within each imputed dataset and then arrive at an overall treatment effect estimate by averaging the effects obtained across imputed data sets (known as within integration). Alternatively, one can average the PSs across the imputed data sets, obtain one PS before estimating treatment effect from PSM (known as across integration).[47–49,53]

We are not the first to consider MI methods when using PSM for causal inference.[47–49] However, significant gaps in methods remain, as work to date has been limited and has consisted of only one form of passive imputation (where confounders are first imputed without consideration of the PS, which is subsequently estimated) along with within and across integration strategies.[47–49] We build upon this excellent body of literature by evaluating active imputations and variations of passive imputations that allow the consideration of auxiliary terms in the imputation model. Further, there is no consensus as to how to best estimate the uncertainty of the treatment effect. This paper presents a novel simulation study to comprehensively evaluate imputation and integration approaches in the context of PSM for the purpose of causal inference. Section 2 details gaps in the current literature that examined MI for PSM. Section 3 describes our methods for conducting a simulation study. We present our findings in Section 4 and discuss interpretation of our findings that inform best statistical practice in Section 5.

## 2. Multiple Imputation Methods for PSM

MI is a simulation based statistical tool to handle missing data, which involves three main steps. In Step 1, multiple sets of plausible values of the missing variable are generated based on the posterior predictive distribution of observed variables to reflect the uncertainties of the imputation process. In Step 2, analyses are performed within each imputed dataset independently, before their results are combined with the application of Rubin's Rules in Step 3.[54] It has been well established in the MI literature that the outcome should always be included in the model when regression parameters are of interest.[38,40,55–57] In the context of PSM, the various strategies we consider (described below) involve Steps 1 and 3.

With respect to Step 1, there are two broad categories of MI strategies that have been introduced in the literature for derived variables or variables that are functions of other variables: active (*MI-active*) and passive (*MI-passive*) (**Figure 1a**). Such derived variables include interaction terms, higher order terms, ratios of two variables (e.g. body mass index), and rates of change.[42–46] In *MI-active*, the derived variable is imputed as if it were any other variable.[42] The simplest, "regular" form of *MI-active*, *MI-regActive*, involves calculating the derived variable in complete cases and imputing it together with all other missing variables in the imputation process, with no consideration of its known relationship to the variables involved in its derivation. *MI-regActive* is a proper imputation method, as all the relationships specified in the scientific model are included in the imputation models, i.e. the imputation model is congenial with the scientific model.[40,50–52] Although *MI-active* is advantageous for its consideration of entire covariance structure, some argue that it undermines the imputation process by creating internally inconsistent values. This motivated a re-derived version of MI-active where the derived variable is recalculated post-imputation (*MI-redActive*).[42]

In contrast to *MI-active* approaches, *MI-passive* approaches maintain the internal consistency between variables used to construct the derived variable and the derived variable itself.[40] In this case, the derived term is not to be imputed but derived after imputing the variables involved in the term's construction. The simplest form of *MI-passive* is *MI-derPassive*, where all variables to be used in deriving the term are imputed using MI, before the term is derived from the imputed data.[42] However, because the derived variable is not included in the imputation process, *MI-derPassive* may be biased, as the process does not consider the entire covariance structure of all variables in the scientific model. Another form of *MI-passive*, *MI-regPassive*, was developed to partially address this issue by including the derived variable in the imputation process of those variables (e.g. an auxiliary variable) that are not involved in its derivation.[58] Auxiliary variables are those that can improve imputation through their inclusion in the imputation process, but do not provide any useful information for the scientific model.[39] For example, a variable that is associated with the pattern of missingness or the missing variable itself can be considered an auxiliary variable.[39] Previous work in MI for PSM has been limited to *MI-derPassive*. Neither *MI-regPassive*, which involves an auxiliary variable, fully or partially observed, nor any active approaches (*MI-regActive* and *MI-redActive*) have been considered previously for handling missingness in PSM.

PS estimates need to be integrated in the analysis to estimate the treatment effect. There has been considerable work in examining integration methods for *MI-derPassive*. Specifically, the PS can be estimated and incorporated within each imputed data set (*INT-within*) prior to obtaining the treatment effect through summarization in Step 3, or the PS can be averaged across the imputed data sets after completing Step 1 and applied to the original data set to obtain the treatment effect (*INT-across*).[47–49] An additional variation on *INT-across*, *INT-across2* has been previously applied in the context of IPTW. It involves averaging both the estimated regression coefficients corresponding to the covariates used to estimate the PS model and the covariates values themselves to arrive at one PS that can be applied to obtain the treatment effect.[53] The rationale is that the PS coefficients are more suitable for combination using Rubin's Rules given their distributional properties than are the PSs themselves, which are confined to be between 0 and 1 (**Figure 1b**).

We comprehensively evaluate the different combinations of MI imputation and integration strategies described.

How to best estimate the variance of the treatment effect in the context of PSM when applying MI is an open research topic.[47–49] In addition to the complications in variance estimation in PSM mentioned above in the absence of missing data, we need to consider the uncertainty introduced by the MI process in the presence of missing data. Rubin's Rules may be applicable to *INT-within*, however, it is unclear how to capture this uncertainty when using *INT-across* and *INT-across2*.[47–49] Bootstrap methods have been proposed in the context of MI[59,60] and specifically with respect to PS-based methods[49,61]. For example, Austin & Small evaluated two potential estimators for PSM in the absence of missing data, where the variance was obtained by either resampling matched pairs or the original observations.[28] Although the former performed well relative to the empirical variance, the latter was comparable and enabled extension to the MI context. Schomaker & Heumann evaluated four bootstrapped-based approaches in the context of MI when PSM was not considered.[59] One of these approaches, "Boot MI", is applicable to the PSM context and overlaps with the ideas described by Austin & Small. We therefore compare two competing variance estimators in this study to better inform those applying MI in the PSM setting: a bootstrapped-based variance estimator and a robust cluster variance estimator (with Rubin's Rules when applying *INT-within*) to account for various sources of variation when MI is applied in the context of PSM.

## 3. Simulation Study Design

We conducted an extensive simulation study to assess the performance of various MI-based strategies and commonly applied methods when estimating the treatment effect using PSM. In all scenarios, we included two binary confounders of the relationship between treatment and outcome, a binary variable representing the treatment or exposure of interest, and a continuous outcome. Missing values were present in one of the two confounders whereas the treatment and outcome variables were always fully observed. For each scenario, 1,000 simulated data sets were generated, each consisting of $n = 2000$ subjects. All data analyses were conducted in R version 3.5.1.[62] MI and PSM were implemented using the *mice* and *Matching* packages respectively.[63,64] The R code to replicate this study is publicly available in a Github repository at https://github.com/yling2019/psm_mi. Below we provide details on the data generation, missing data mechanisms, missing data methods considered, and metrics for performance evaluation.

### 3.1 Data generation

*Confounders.* Two binary variables $X = (X_1, X_2)$ that confound the relationship between treatment and outcome were generated, by first creating two variables from a bivariate normal distribution (correlation of 0.5) each with mean 0 and variance 1, which were then dichotomized at the mean.

*Treatment indicator.* A binary treatment variable $T$ was generated from a binomial distribution such that:

$$logit\,(p\,(T = 1|\,X_1, X_2)) \;=\; \alpha_0 \,+\, \alpha_1 X_1 \,+\, \alpha_2 X_2$$

where $\alpha_1 = \alpha_2 = 2$ so both covariates contributed equally to the treatment assignment. The intercept of the treatment $\alpha_0$ was selected such that roughly 30% of subjects were treated, to reflect real-world datasets where there are often many more control subjects than treated.

*Outcome.* A continuous outcome variable $Y$ was generated as a linear function of the treatment and both covariates.

$$Y \;=\; \beta_0 \,+\, \beta_1 X_1 \,+\, \beta_2 X_2 \,+\, \beta_t T \,+\, \varepsilon$$

where $\beta_1 = \beta_2 = 2$ so both covariates were equally and positively associated with outcome and $\varepsilon \sim N(0, 10^2)$. The intercept $\beta_0$ was set to zero and the true treatment effect $\beta_t$ was set to 2.

*Auxiliary variables.* An auxiliary variable $Z_2$ was generated to be highly correlated with $X_2$ (correlation = 0.98). An additional auxiliary variable $Z_{ps}$ was generated to be highly correlated with the estimated PS score (based on full observed data, with correlation of 0.98). More specifically, setting $\delta_{02} = 1, \delta_{0ps} = 0, \delta_{12} = \delta_{1ps} = 10$, the auxiliary variables were generated as:

$$Z_2 = \delta_{02} + \delta_{12}X_2 + \varepsilon$$

$$Z_{ps} = \delta_{0ps} + \delta_{1ps}\widehat{PS} + \varepsilon$$

where $\varepsilon \sim N(0, 1^2)$.

**3.2 Missing data mechanisms (MDMs)**

Missingness was always induced in $X_2$, whereas $X_1$ was fully observed. We induced missing data in $X_2$ according to five different mechanisms: MCAR, MAR1, MAR2A, MAR2B and an MNAR scenario. Whereas MAR1 represented a simple MAR scenario, MAR2A and MAR2B were more sinister scenarios that captured the complexity of MDM in real-world datasets. In MAR1, MAR2A, and MNAR, missingness was related to treatment and outcome. In MAR2A, missingness was also related to $Z_2$, the auxiliary variable associated with $X_2$. In MNAR, missingness was also related to $X_2$. Let $R_2$ be an indicator variable denoting whether $X_2$ is missing ($R_2 = 1$) or not ($R_2 = 0$). We set $Y_b$ and $Z_{2b}$ to be dichotomizations at the median of the outcome variable Y and auxiliary variable $Z_2$ respectively. Under each MDM, the intercept $\gamma_0$ was selected such that 50% of the observations were missing. Let $\gamma_{11} = 5, and\ \gamma_{00} = 1$, and $I$ be an indicator variable. Missingness in $X_2$ was induced as follows:

$$\text{MAR1: } logit(R_2 = 1) = \gamma_0 + \gamma_{11}I_{t=1,yb=1} + \gamma_{00}I_{t=0,yb=0}$$

$$\text{MAR2A: } logit(R_2 = 1) = \gamma_0 + \gamma_{11}I_{t=1,yb=1,z2b=1} + \gamma_{00}I_{t=0,yb=0,z2b=0}$$

$$\text{MNAR: } logit(R_2 = 1) = \gamma_0 + \gamma_{11}I_{t=1,yb=1,x2=1} + \gamma_{00}I_{t=0,yb=0,x2=0}$$

To study the impact of having a partially observed auxiliary variable, we also induced missingness in $X_2$ according to a second MAR2 missing mechanism, MAR2B, based on treatment, outcome, and PS. Letting $Z_{psb}$ be the dichotomizations of $Z_{ps}$, missingness in $X_2$ was induced as follow:

$$\text{MAR2B: } logit(R_2 = 1) = \gamma_0 + \gamma_{11} I_{t=1, yb=1, zpsb=1} + \gamma_{00} I_{t=0, yb=0, zpsb=0}$$

Additionally, we induced missingness in the auxiliary variable, $Z_{ps}$, under three scenarios that assumed MAR2B for $X_2$: aux_MCAR, aux_MAR1, and aux_MAR2. In both aux_MAR1 and aux_MAR2, missingness was related to $T$ and $PS_b$, where $PS_b$ is the dichotomization at the median of the PS estimated using full data prior to inducing missingness. Let $R_z$ be an indicator variable denoting whether $Z_{ps}$ is missing ($R_z = 1$) or not ($R_z = 0$). The intercept term $\varepsilon_0$ was selected to ensure 20% missingness in $Z_{ps}$ and missingness can be expressed as:

$$logit(R_z = 1) = \varepsilon_0 + \varepsilon_{11} I_{t=1, psb=1} + \varepsilon_{10} I_{t=1, psb=0} + \varepsilon_{01} I_{t=0, psb=1} + \varepsilon_{00} I_{t=0, psb=0}$$

where $\varepsilon_{11} = 5, \varepsilon_{10} = 0, \varepsilon_{01} = 0, \varepsilon_{00} = 5$ in aux_MAR1 and $\varepsilon_{11} = 0, \varepsilon_{10} = 5, \varepsilon_{01} = 5, \varepsilon_{00} = 0$ in aux_MAR2.

### 3.3 Missing data methods

*Common missing data methods.* We applied various missing data methods that are widely used in the medical research literature including CC, CVA, mean imputation, and the use of missing data indicators.

*Multiple imputation strategies.* **Figure 1** displays the MI strategies considered for PS estimation and integration. *MI-derPassive*, *MI-regActive*, and *MI-redActive* were applied under MCAR, MAR1, MAR2A, and MNAR conditions with or without auxiliary variable $Z_2$ in the imputation model, where $Z_2$ had no missing values (**Table 1**). Under the MAR2B MDM, we included an additional partially observed auxiliary variable $Z_{ps}$ in the imputation model when *MI-regPassive* and *MI-derPassive* were applied. Under this scenario we also examined performance by order of inclusion of the variables in the imputation model (i.e., whether $X_2$ was imputed before $Z_{ps}$ or not). Integration approaches considered were *INT-within*, *INT-across*, and *INT-across2*. Note that *INT-across2* cannot be combined with *MI-regActive*, as the PS was directly imputed from MI in *MI-regActive*. In MICE, 50 multiply imputed datasets ($m = 50$)[43], five iterations ($maxit = 5$) and default settings for imputation method (predictive mean matching for continuous variable and logistic regression for binary) were used. The outcome was included in all imputation models.[40]

### 3.4 PSM and treatment effect estimation

We estimated coefficients $\alpha_1$ and $\alpha_2$ using a correctly specified logistic regression model, $logit\ (p(T = 1\,|\,X\,)) = \alpha_0 + \alpha_1\,X_1 + \alpha_2 X_2$. PS scores were estimated as the fitted values of the regression model on the response scale. One-to-one nearest neighbor matching without replacement was applied. Subjects were matched by PS scores with calipers of width that is 0.2 of the standard deviation of the logit of the propensity score.[16,65] After matching subjects, the treatment effect was estimated using standard linear regression methods, by regressing $Y$ on $T$ and confounders $X_1$ and $X_2$ to obtain the estimate for the beta coefficient representing $T$[26] with the exception of cases where *INT-across* was applied, as in the presence of multiple data sets, there were multiple sets of $X_1$ and $X_2$.

**3.5 Variance estimation**

In the absence of missing data, we used two approaches to estimate the uncertainty of the treatment effect: (1) a robust cluster variance estimate[66] that accounts for the matched design and (2) a bootstrapped variance calculated as the standard deviation of treatment effects in 1,000 bootstrapped samples to account for both PS estimation and the matching process. When commonly applied missing data methods were considered, the robust cluster variance estimator was used. When MI was applied in the context of PSM, we compared 1) the robust cluster variance estimator and 2) a bootstrapped variance. For the former, when the integration strategy was *INT-within*, a robust cluster variance was estimated within each of the $m$ imputed dataset, before application of Rubin's Rules to yield one final variance. For both *INT-across* and *INT-across2*, Rubin's Rules do not apply; instead we obtain only one robust cluster variance. For the latter, the detailed procedure is described as follows:

1. Sample with replacement $n = 2000$ rows from the observed dataset $D = (X, T, Y, Z, R)$ to obtain a bootstrapped dataset $D_b$ which contains missing values;

2. Impute $m$ datasets for $D_b$ using the imputation strategy (*MI-derPassive, MI-regPassive, MI-regActive,* or *MI-redActive*), for $k = 1,2,\ldots,m$, denoted as $D_b(k)$;

3. Apply the integration approach (*INT-within, INT-across* or *INT-across2*) to obtain a single effect estimate for $D_b$;

4. Repeat steps 1-3 $B$ times to obtain $B$ bootstrap replicates from which treatment effect $\beta_{tb}$ can be estimated for a given bootstrap sample $D_b$;

5. Calculate bootstrapped standard error as the standard deviation of B treatment effects estimated from each bootstrap sample: $SE_{bootstrap} = sd(\beta_{tb}) \ for \ b = 1,2,\ldots,B$

## 3.6 Performance metrics

After PSM, we examined the percentage of treated subjects matched and the standardized differences of covariates after matching. For mean imputation, standardized differences were calculated in the original full data and the imputed data. For missing indicator variables, standardized differences were calculated in the full data without missingness, as well as its observed and missing part. For *INT-within,* standardized differences were calculated in 1) each of the imputed datasets 2) the full dataset. For *INT-across* and *INT-across2*, standardized differences were calculated in 1) the average of $m$ imputed dataset and its observed and imputed parts respectively 2) the full dataset.[53,55] For each missing data method, we report on bias, variance, mean squared error (MSE), relative MSE (relative to PSM in the full dataset), and coverage probability summarized over 1,000 simulations per scenario. The robust cluster variance and bootstrapped variance were compared to their corresponding empirical variance for each MI strategy. Coverage was estimated as the proportion of 1,000 simulations such that the interval $[\widehat{\beta}_t - 1.96 \times SE, \widehat{\beta}_t + 1.96 \times SE]$ contained the true treatment effect of $\beta_t = 2$ ($SE$: robust cluster standard error or bootstrapped standard error). We used the normal theory estimator because the percentile based method didn't not perform well in simulations by Austin & Small[28], and calculating accelerated and bias-corrected confidence intervals (BCa)[31] is too computationally intensive. Monte Carlo standard errors were calculated for bias, empirical standard error, MSE, and coverage.[67] Reference metrics for missing dat methods were based on applying PSM to the full data (PSM_full).

## 4. Results

We validated the data generation process by comparing the resulting bias and standard error from fitting the true data generating model to that obtained from applying PSM to the full data (PSM_full). Both methods yielded unbiased treatment effect estimates (bias = -0.006 in both cases). PSM yielded a higher standard error as expected due to discarding unmatched samples (0.313 using regression in the full dataset and 0.380 using PSM_full). Coverage reached the nominal level of 95% using both methods. These results matched well with their corresponding empirical standard error (0.306 and 0.376 respectively). In PSM, the robust cluster standard error and bootstrapped estimators were comparable (0.380 in both cases) and close to the empirical (0.376).

### 4.1 Commonly applied missing data methods

Of the commonly applied approaches, CC had the most favorable MSE relative to that the of PSM_full (rMSE = 1.857 to 48.658, **Appendix Table A1**). CC produced biased treatment estimates (bias = -2.489, -0.815, and -1.084 in MAR1, MAR2A, and MNAR respectively) and less efficient estimates relative to PSM_full (robust standard error =0.537, 0.838, 0.682, 0.682 in MCAR, MAR1, MAR2A and MNAR respectively vs 0.380 in PSM_full). CVA, mean imputation, and the use of missing indicators yielded greater bias relative to CC (5.058 to 5.059 for CVA; 2.985 to 5.759 for mean imputation; and 2.973 to 5.534 for missing indicator), although their robust cluster variances were smaller than that of CC. Comparisons of statistical properties obtained when not adjusting for $X_1$ and $X_2$ were similar.

**4.2 Variance estimation in MI-based strategies**

While the robust cluster variance estimator and the bootstrapped-based variance estimator were comparable in the absence of missing data, differences were observed in the presence of missingness and when MI was applied. Specifically, the robust cluster variance estimator consistently underestimated the empirical variance in *INT-across* approaches. Among the *INT-across* approaches, the ratio of the robust variance to the empirical exceeded 0.8 only when an auxiliary variable was included (**Figure 2**). The worst performance for the robust estimator was observed in *MI-regActive INT-across* approaches, where the variance ratios were lower than 0.1. In contrast, the variance was consistently overestimated in *INT-within* approaches, where the variance ratio surpassed 5 under *MI-regActive* and *MI-redActive* approaches, especially when an auxiliary term was used. The ratio of the robust estimator for the variance relative to the empirical under *INT-across2* was close to 1 across all MI methods and MDMs. On the other hand, the bootstrapped-based variance was more comparable to the empirical variance across all MI integration strategies; the ratio of bootstrapped variance to the empirical ranged from 0.675 to 1.875, with mean 1.01. We did not observe any trend specific to imputation methods, integration methods, or the inclusion of auxiliary variable in the imputation model. All subsequent results were therefore calculated using the bootstrapped-based variance estimator.

**4.3 Comparing various MI strategies**

For simplicity, we start by describing performance of MI strategies under MAR2A, as MAR2A was induced to reflect a realistic but complex MDM. For all MI strategies, balance was achieved such that the absolute standardized difference in $X_1$ and $X_2$ between treated and controls based on the imputed dataset was below 0.1 with the exception of *MI-regActive* in the presence of an auxiliary term (**Appendix Table A2**). Performance between *MI-derPassive INT-across* and *MI-redActive INT-across2* was comparable; both were top performers when considering both bias and efficiency. Specifically, *MI-derPassive across* achieved the lowest rMSE (10.073), followed by *MI-redActive INT-across2* (10.435) and *MI-derPassive INT-across2* (11.717) (**Table 2** and **Figure 3**). Performance under *MI-derPassive* and *MI-regActive* strategies was consistent across integration approaches, with the former identified as having among the best statistical properties and the latter as performing poorly with respect to both bias and efficiency. In contrast, the performance of *MI-redActive* varied by integration approach, where *MI-redActive INT-within* was the worst performer (rMSE=87.819, bias= 3.433) and *MI-redActive INT-across2* was the second-best performer (rMSE=10.435, bias= -0.950). INT-across2, INT-

*across*, *INT-within* were ranked from the lowest to the highest with respect to rMSE, which was largely driven by the bias.

**4.4 The impact of auxiliary terms**

Under MAR2A, when a fully observed auxiliary term, $Z_2$, was included in the imputation model, considerable variation was observed. *MI-derPassive INT-across* (rMSE= 1.027, bias = 0.039), *MI-redActive INT-across* (rMSE= 1.056, bias=0.045), and *MI-derPassive INT-within* (rMSE = 1.672, bias = 0.319) were the top three performing MI strategies in terms of rMSE (**Table 2** and **Figure 3**). Interestingly, including an auxiliary term improved the rMSE for all strategies, except for *MI-redActive INT-across2* and *MI-derPassive INT-across2*, the top two performing strategies when the auxiliary variable was not included. The auxiliary term improved efficiency for most MI strategies except for *MI-regActive INT-across*. The auxiliary term improved the absolute bias for most MI strategies except for *MI-redActive INT-across2* and *MI-derPassive INT-across2*. Inclusion of the auxiliary variable was required to obtain nominal level of coverage probability in both MAR2A and MNAR (**Figure 4**).

The performance of MI strategies is shown in **Appendix Table A4** and **Figure 5** under a modified MAR2 scenario (MAR2B) where missingness was a function of a different auxiliary variable, $Z_{ps}$, treatment, and outcome. For reference, we first evaluated performance of passive approaches when $Z_{ps}$ was fully observed. *MI-derPassive INT-within* achieved the lowest rMSE and bias (rMSE=1.032, bias=-0.019), followed by *INT-across* (rMSE=1.394, bias=-0.205), and *INT-across2* (rMSE= 10.319, bias=-1.016). The bootstrapped standard error was the largest in INT-across2

(0.677) and comparable for INT-within (0.378) and INT-across (0.380). When $Z_{ps}$ was partially observed, *MI-derPassive* and *MI-regPassive* were largely comparable. Although *INT-across* methods yielded the smallest bootstrap standard error, *INT-within* methods resulted in smaller bias and MSE. While imputing $X_2$ before or after $Z_{ps}$ affected the resulting bias, bootstrap standard error and MSE, the order did not change our conclusions on the best performing imputation and integration MI strategies as mentioned above.

**4.5 Comparison across MDMs**

When auxiliary variables were not incorporated into the imputation model, *MI-derPassive* outperformed *MI-redActive*, followed by *MI-regActive* in terms of rMSE, bias, and efficiency (**Appendix Table A3** and **Figure 3, Figure 4**). While *MI-derPassive INT-within* demonstrated strong performance regardless of the presence of an auxiliary term under MCAR and MAR1, *MI-derPassive INT-across* outperformed *INT-within* under MAR2A and MNAR, especially when an auxiliary variable was included in the imputation model. Inclusion of an auxiliary variable did not greatly improve properties of the top performers under MCAR and MAR1. In contrast, the performance of MI-strategies was much improved under MAR2A and MNAR, where missingness was related to the auxiliary variable.

**5. Discussion**

We investigated several pragmatic research questions concerning how to optimally apply MI when utilizing PSM in the presence of a partially observed confounder. We compared the performance of non-MI missing data methods that are commonly applied along with various MI-based strategies that vary both in how the PS is estimated or imputed and in how the PS is integrated into the analysis. In addition, we evaluated the impact of inclusion of an auxiliary term in the imputation model on the ranked performance of the strategies as well as the impact of the order of inclusion when there is more than one variable with missing data. Among the commonly applied missing data methods, CVA and single imputation methods (mean imputation and missing indicator imputation) led to large bias in our simulation study. In contrast, CC was not as biased due to the use of a caliper that ensured only those subjects with closely matched PSs were included. CC did, however, suffer from loss of efficiency. There was large heterogeneity among the MI strategies considered. While *MI-derPassive INT-within* performed well in MCAR and MAR1, inclusion of the auxiliary variable in the imputation model was necessary to achieve nominal coverage under MAR2A and MNAR. Based on our results, we recommend applied researchers to 1) adopt *MI-derPassive* approaches; and 2) consider *INT-within* for mild MDMs (MCAR and MAR1) and *INT-across* for MDMs that are more complex. We note that this choice requires deep thought into the MDM and relies on unverifiable assumptions and our understanding of the nature of the study and variables involved. Sensitivity analyses that demonstrate no difference may be reassuring, whereas those that highlight differences may reflect incorrect assumptions about the MDM. We also recommend 3) use of the bootstrap to estimate variance; and 4) inclusion of key auxiliary variables in the imputation model if available.

Our study is important in identifying the limitations of commonly applied methods. Considerable bias and inefficiency were observed among all commonly applied methods relative to that yielded by applying PSM to the data set with no missing values. At least one of the MI-based strategies always outperformed the commonly applied methods. Importantly, when applying PSM, it is well established that balancing diagnostics are useful tools to guide analyses, and this proves difficult with commonly applied methods. In particular, if balance -- as reflected by standardized differences in covariates -- is not achieved, additional differential modeling of the PS should be considered. It is important to note, however, that in the presence of missing data, simple diagnostics are not straightforward to obtain. For example, while CC can be applied using those matched pairs where balance is achieved, bias may still occur because of the observations not included due to missingness. For missing indicators, we were unable to achieve balance in the variable with missing data, as revealed in our simulations when we parsed the data by observed and missing data to evaluate balance in these respective parts. Thus, in practice, one may have a false sense of the balance as the user is only privy to assessing the balance in the observed data. Similarly, application of mean imputation distorts the distributional properties of the variable with missing data, potentially yielding a distorted view of balance when the imputed values are utilized in calculating the standardized differences (**Appendix Table A1**). Finally, for CVA, a false sense of security may be given when evaluating balance in only one variable, when exclusion of the other variable could lead to bias.

We have primarily examined the differences between passive and active MI methods when a derived variable, such as PS, was considered in the analysis and only partially observed. *MI-*

*derPassive* methods surpassed *MI-regActive* approaches in almost all performance metrics across all MDMs. Although active MI methods were proposed so the entire covariance structure of all variables including the PS itself could be retained, the covariance structure between PS and the PS variables is complex and difficult to learn using complete cases only. Unlike usual derived measures (e.g. interactions and higher order terms) that are derived as a simple function of other variables, PSM requires estimation. In this way, PS is different from other derived variables in that the exact function will vary depending on the data considered. This was particularly relevant for MDMs that were not MCAR, where PS -- estimated from complete cases may be biased depending on the MDM – introducing further bias into the imputation procedure and consequently the treatment effect. Such bias was also reflected in the difficulty to achieve balance both in the imputed and fully observed data under *MI-regActive* (**Appendix Table A2**). Further, the poor estimation of the treatment effect has implications for estimates of uncertainty. More specifically, the bootstrapped-based variance involves multiple draws of the data with application of *MI-regActive* coupled with PSM performed on each draw. As the bias introduced in the estimation of the treatment effect highly varies across the draws, this leads to an increased estimate of the variation (**Appendix Figure A1**). Although adding an auxiliary variable reduced bias for most of the cases, it did not help reduce bias for the extreme cases where the bias without auxiliary variable was unexpectedly high. Nevertheless, this problem might not be as profound in real data analyses where there a larger number of covariates are typically included in the PS model. In contrast, *MI-redActive* can be thought of as a hybrid between *MI-derPassive* and *MI-regActive* and mitigated this issue by re-deriving PS post-imputation, as in *MI-derPassive*. Although *MI-derPassive* still outperformed or performed comparably to *MI-redActive* methods in our simulations, our findings demonstrated that *MI-active* approaches hold great promise in recovering the underlying covariance structure especially when there are higher dimensional data and more flexible models are being considered.

Although we found *MI-derPassive* to be the optimal MI method for PSM with partially observed covariates, there are several considerations including the integration choice. While Mitra & Reiter recommended *INT-across,* de Vries & Groenwold found *INT-within* to yield estimators with better statistical properties.[48,49,57] In a different setting (IPTW instead of PSM), Leyrat et. al. demonstrated superior properties of *INT-within* methods over *INT-across* and *INT-across2*.[53] Our results varied across MDMs and were consistent with Leyrat and others and de Vries & Groenwold under MCAR and MAR1 but corroborated with those of Mitra and Reiter under MAR2A and MNAR conditions. We also share the perspective of Leyrat and others[53] that it was more straightforward to assess balance in *INT-within* strategies and observed that the covariates were mostly balanced in both the imputed and full data sets (**Appendix Table A2**). Further, Leyrat and others pointed out that the *INT-across* and *INT-across2* produced consistent estimators only when both the observed and imputed data were balanced.[53] We therefore paid close attention to balance diagnostics for *INT-across* and *INT-across2* methods but did not observe balance in both parts of the data (observed and missing) under all MDMs other than under MCAR.

An important contribution of our paper is resolution of how to estimate the variance when doing PSM and applying MI. There has been extensive but conflicting research on this topic in the context of MI and IPTW where Rubin's Rules have been recommended for *INT-within* by some authors[53,68] and a bootstrapped-based estimator was recommended by others[61]. Relative to IPTW, however, we are faced with the additional issue of capturing the uncertainty of matching in PSM. Prior studies of MI applications in the context of PSM acknowledged this issue[47,48], but only one study has explicitly stated their recommendation of a bootstrapped-based variance[49], although the choice was not studied comparatively or discussed fully. In our study, we found that application of Rubin's Rules when the robust cluster estimator was used for each imputed data set overestimated the variance under *INT-within* approaches and underestimated it under *INT-across* approaches. We therefore agree with de Vries & Groenwold[49] in recommending the bootstrapped variance, as it captures the uncertainty of PS estimation, matching procedure, and imputation process. Further, it demonstrated good performance with respect to the empirical variance. We acknowledge the lack of theoretical support for this choice, which comes with challenges, as the estimator for the treatment effect based on PSM and MI is not a smooth function. Although Abadie & Imbens[24] proved that the bootstrap variance is not valid in matching *with* replacement, their results may not be applicable in our study when matching was done *without* replacement, where one control unit can only be used for matching at most once[28]. Other alternative non-parametric solutions with stronger theoretical justification, such as subsampling, has their own limitations (e.g. the need for a sufficiently large sample size and a burden on the user to appropriately select a sub-sample and replication size).[69]

Auxiliary variables are often useful for adhering to a MAR MDM, but not always possible in the context of PSM. Specifically, variables related to partially observed confounders may be considered confounders themselves and thus, may not exist outside of estimation of the PS. Our team has worked on studies, however, where auxiliary terms may be available. For example, in a comparative effectiveness study of anticoagulants among kidney transplant patients, a PS that balances patient characteristics may include body mass index (BMI) at treatment initiation but not BMI at transplant listing. The latter is an excellent candidate for an auxiliary variable that can aid in imputing BMI at treatment initiation as well as other PS covariates. By including a strong auxiliary variable in the imputation process, we showcased the maximal performance improvement given any auxiliary variable. In practice, the strength of auxiliary variable varies and consequently the improvement in performance may be moderate.

There are several limitations to our study. As with any simulation study, we recognize that the limited scope of our simulations may compromise generalizability. Specifically, only two covariates were included in the PS, and there were only main effects in the data generating mechanism. We also only considered the scenario when one confounder was partially observed, whereas missingness of covariates that are not confounders, treatment or outcome was not considered.[47,49,68] Whether our findings extend to binary or time-to-event outcomes remains to be studied, since odds ratios and hazard ratios are not collapsible.[70] Nevertheless, some of our findings agree with those from simulation studies on binary outcomes in a similar context (IPTW).[53,68] Finally, we did not explore the impact of mis-specifying the correct PS model or having completely missing covariates.

Overall, we have addressed an important topic – how to apply MI strategies in the presence of missing values in confounders in the context of PSM. Our work will facilitate future applied researchers' choice of optimal missing data methods in all kinds of statistical analyses that involve PSM. In addition to classical causal inference settings, our results are applicable to other types of studies that utilize PSM including those that generalize randomized clinical trial findings to real-world target populations captured in observational databases.[71,72]

## 6. Acknowledgements

This work was supported by a Sanofi iDEA Award. We are particularly grateful for the excellent insight and guidance on this research provided by the Sanofi research team members: Robert LoCasale, Karen Chandross, Liz Zhou and Cliona Molony.

**Figures**

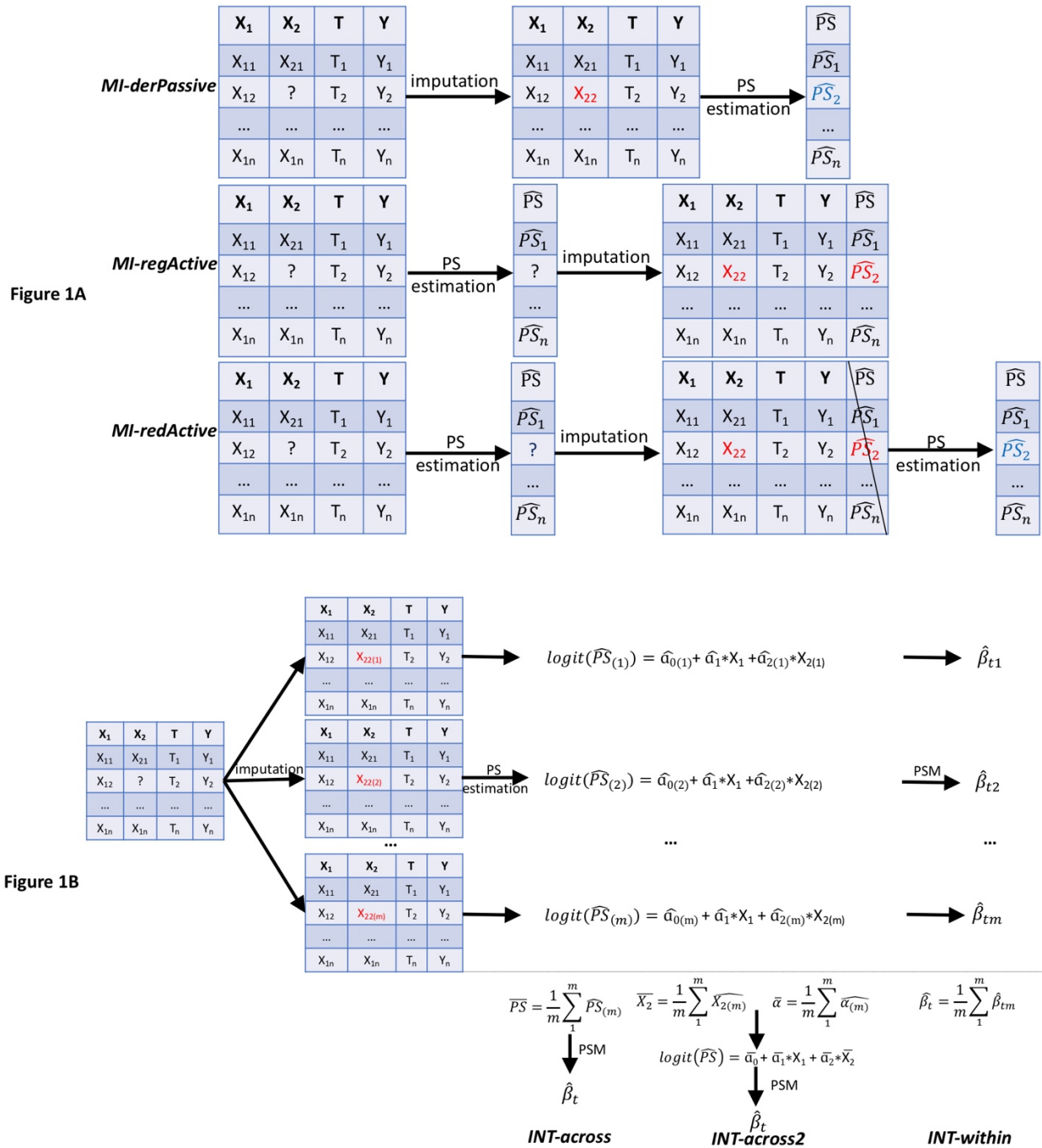

**Figure 1.** Illustration of various multiple imputation (MI) imputation and integration strategies in the context of our simulation study. Figure 1(a) above details three imputation strategies: *MI-derPassive, MI-regActive, and MI-redActive*. Text in red indicates the values that are imputed, while text in blue indicates the values that are derived. *MI-redActive* shares the same two steps with *MI-regActive*, but it discards the PS values output from MI and re-derives PS post-imputation. Figure 1(b) below illustrates three integration strategies: *INT-across,* and *INT-*

*across2, INT-within.* Imputed values are colored in red. For INT-across, PSM analysis was conducted after PSs are averaged over $m$ imputed datasets. For INT-across2, PS was estimated using the average PS coefficients and average covariates followed by PSM. Note that *MI-regActive* cannot be applied in combination with *INT-across2.* For simplicity and without loss of generality, the above illustration does not involve inclusion of auxiliary variables in the imputation model.

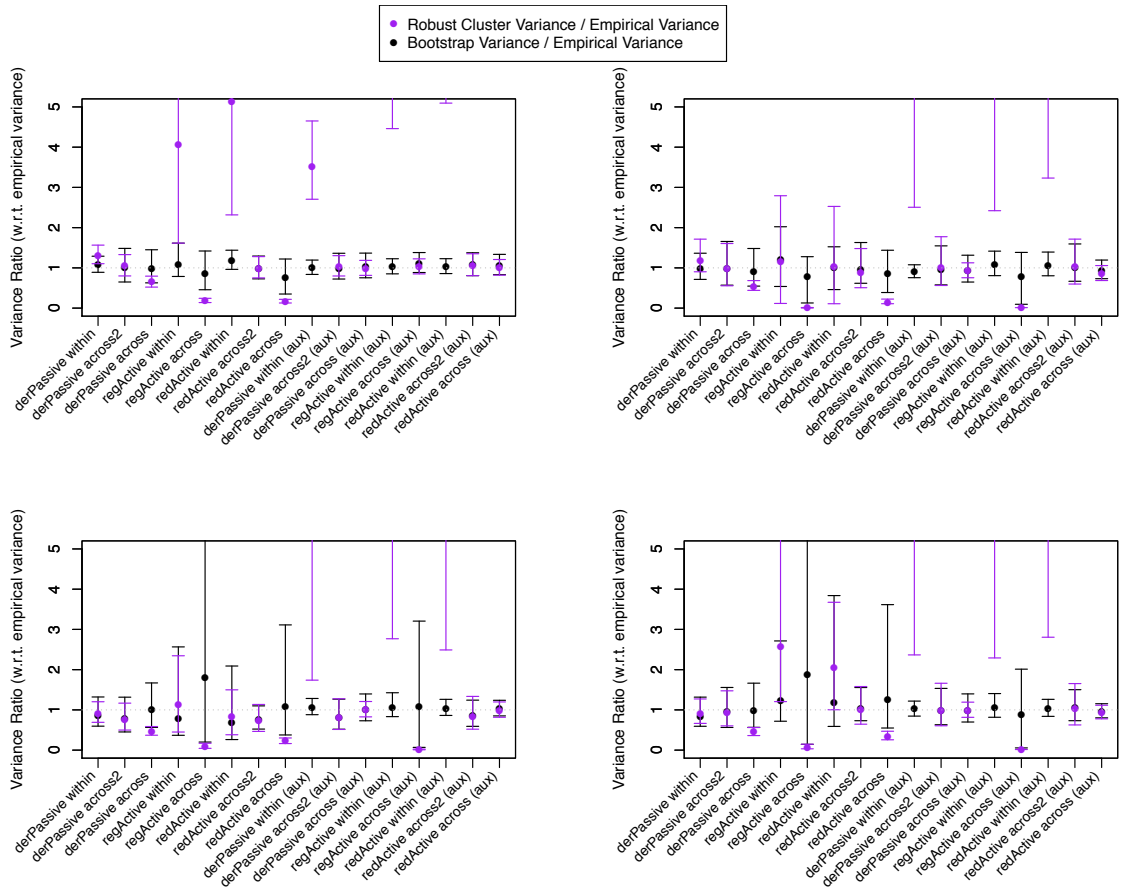

**Figure 2.** Comparing robust cluster variance and bootstrap variance to empirical variance in various MI strategies in all four missing data mechanisms. Variance ratios were calculated by dividing the two variance estimates with the corresponding empirical variance of each specific MI strategy varying imputation methods, integration strategies, with or without auxiliary variable in the imputation model. The 95% confidence interval estimated using the 2.5% and 97.5% percentile of the variance estimated from 1,000 simulations are illustrated for each MI method as well. Note that some of the variance ratios above 5 are not shown.

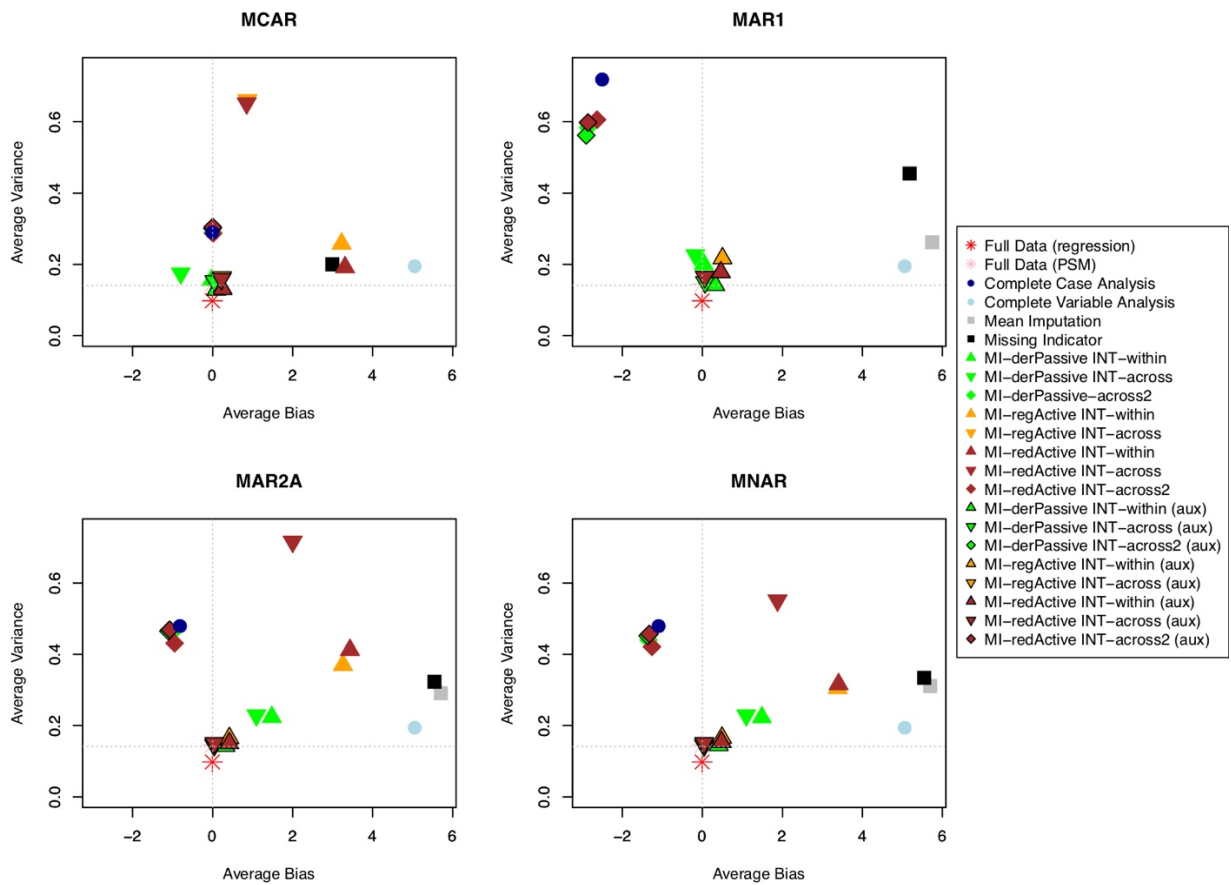

**Figure 3.** Average bias and variance estimated from 1) full data without missingness 2) commonly applied missing data methods including complete case analysis, complete variable analysis, mean imputation and missing indicator 3) various multiple imputation (MI) strategies paired with integration strategies, averaged from our 1,000 simulations in our four missing data mechanisms. Results of methods with variance larger than 1 is not shown on the plot but are recorded in **Table 2** and **Table A2**. For 1) and 2), the robust cluster variance was plotted as this is what applied researchers would use in practice. For 3) the bootstrapped variance was plotted as this is our recommended variance estimator. The vertical and horizontal grey dotted lines indicate the average bias and variance obtained by applying PSM to the full data.

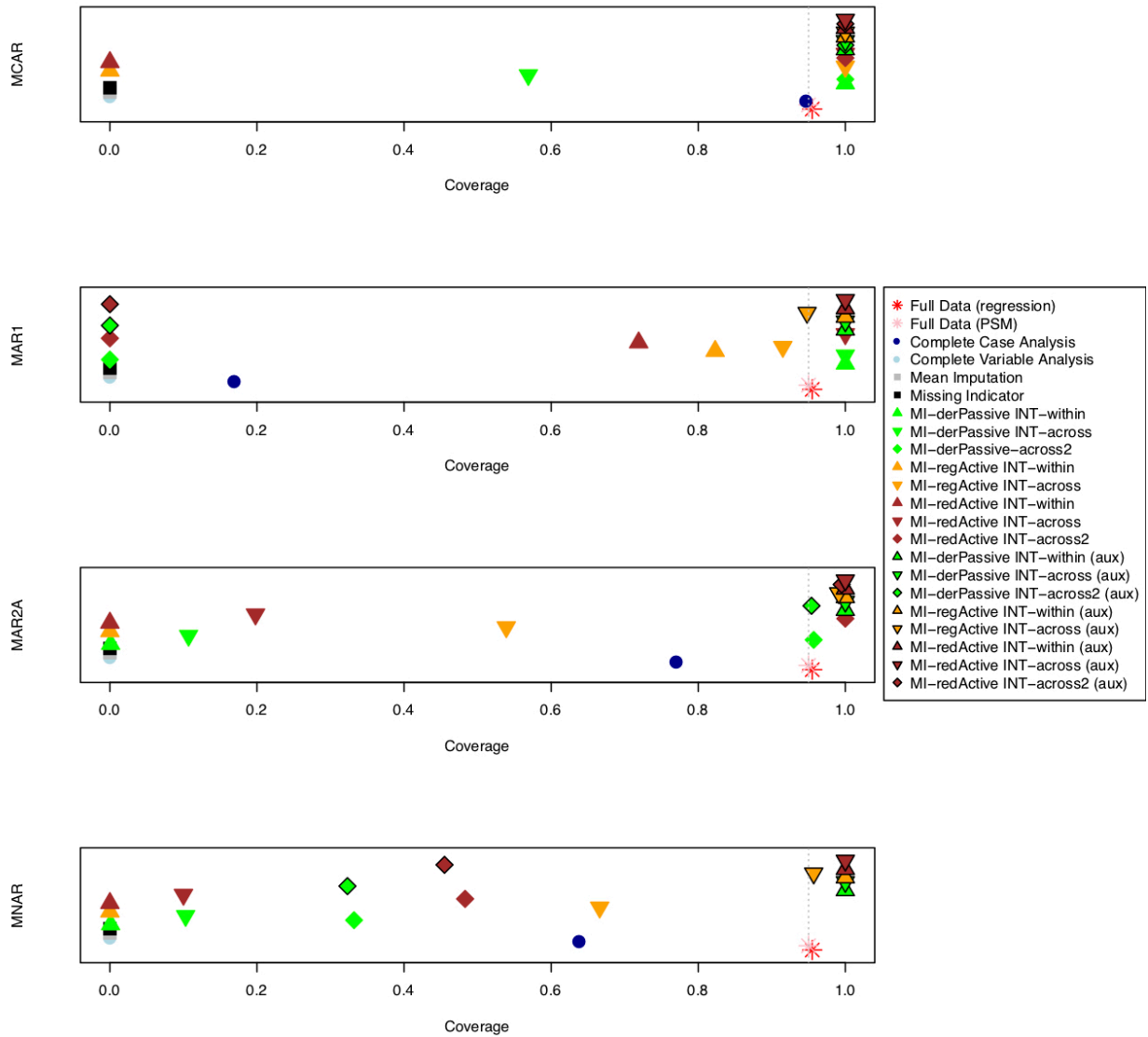

**Figure 4.** Coverage probability of 1) full data without missingness 2) commonly applied missing data methods including complete case analysis, complete variable analysis, mean imputation and missing indicator 3) various multiple imputation strategies paired with integration strategies from our 1,000 simulations in our four missing data mechanisms. For 1) and 2) the coverage was calculated using average bias and robust cluster variance and for 3) the coverage was calculated using average bias and bootstrapped variance. The grey dotted line is at coverage = 0.95.

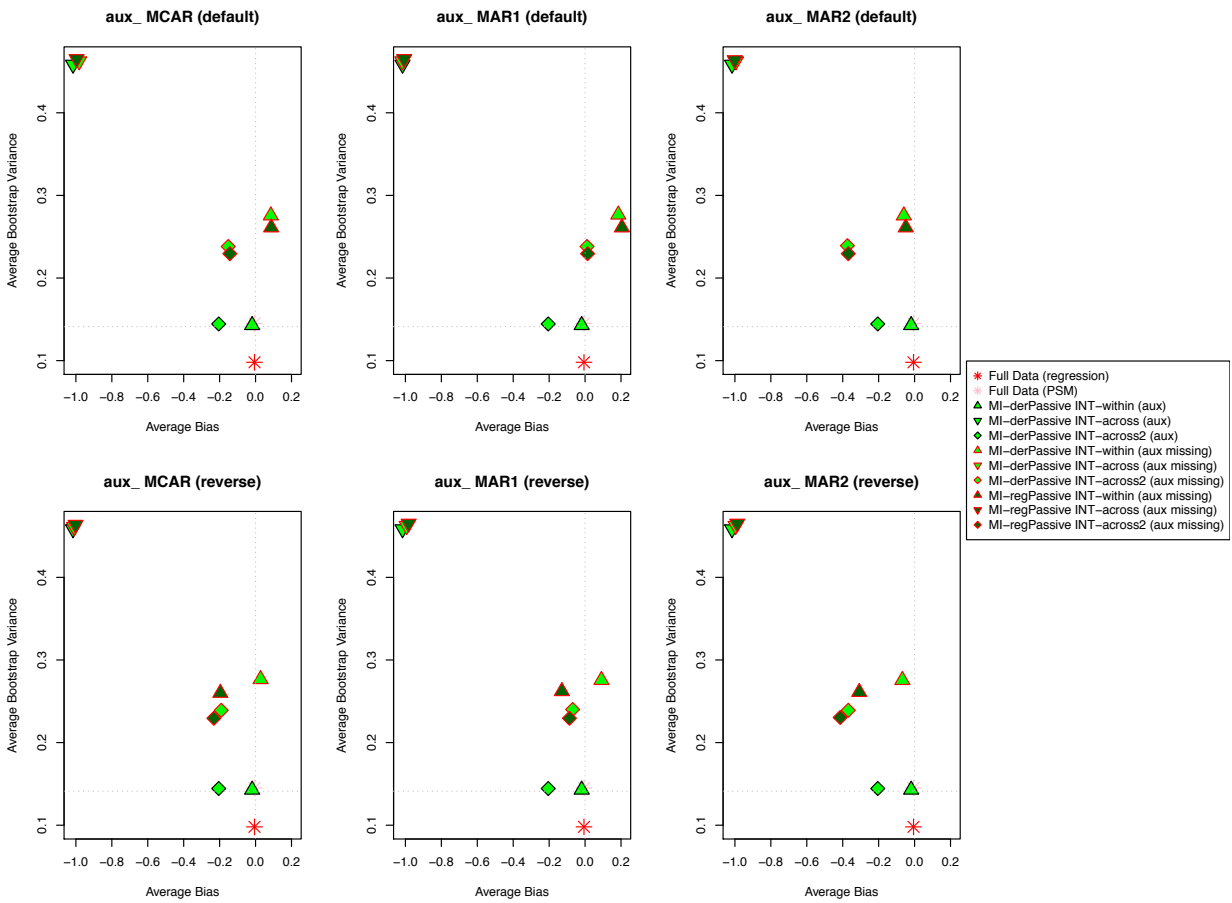

**Figure 5.** Average bias and bootstrap variance estimated when $X_2$ was missing MAR2B (when the missingness of $X_2$ was related to treatment, outcome, and $Z_{ps}$ as described in the Methods section) and *MI-derPassive* and *MI-regPassive* were applied, averaged from out 1,000 simulations across three missing data mechanisms for auxiliary variable $Z_{ps}$. The top three plots show results when $X_2$ was imputed before $Z_{ps}$ in the imputation process, and such order is reversed in the bottom three plots. The vertical and horizontal grey dotted lines indicate the average bias and variance obtained by applying PSM to the full data.

## Tables
## Table 1
Comparison of the different predictors included in the imputation model for the multiple imputation approaches included

| | Scientific model: $Y \sim T + X_1 + X_2$ | | | |
|---|---|---|---|---|
| Missing | Predictors in imputation model | | | |
| Variable | MI-regActive | MI-redActive* | MI-derPassive | MI-regPassive*** |
| $X_2$ | $X_1, T, Y, PS\ (Z_2)$ | $X_1, T, Y, PS\ (Z_2)$ | $X_1, T, Y, (Z_2\ or\ Z_{ps})$ | $X_1, T, Y, Z_{ps}$ |
| $PS$ | $X_1, X_2, T, Y\ (Z_2)$ | $X_1, X_2, T, Y (Z_2)$ | - | - |
| $Z$** | - | - | $X_1, X_2, T, Y$ | $X_1, X_2, T, Y, PS$ |

*$PS$ is re-derived from $X_1$ and $X_2$ after MI procedure

** $Z_2$ was used in *MI-derPassive*, *MI-regActive*, and *MI-redActive* when it was fully observed. $Z_{ps}$ was used in *MI-derPassive* and *MI-regPassive* when 20% of values were missing. Under MAR2B, the order of imputation (whether $Z_{ps}$ was imputed before or after $X_2$) was also evaluated.

*** The steps within MICE algorithm for *MI-regPassive* are described as follows (when $X_2$ was imputed before $Z_{ps}$): 1) impute $X_2$, with $X_1, T, Y$ and $Z_{ps}$ 2) derive $PS$ from $X_1$ and $X_2$ 3) impute $Z_{ps}$ with $X_1, X_2, T, Y$, and $PS$.

**Table 2**. Bias, standard error mean squared error (MSE), relative mean squared error (rMSE), and coverage results of various multiple imputation strategies under MAR2A (Monte Carlo standard errors in parentheses).

| MI Strategy | | | Standard Error | | | | |
|---|---|---|---|---|---|---|---|
| Imputation | Integration | Bias | Empirical | Bootstrap | MSE | rMSE | Coverage |
| *Auxiliary variable not included in imputation model* | | | | | | | |
| MI-derPassive | INT-within | 1.479 (0.016) | 0.511 (0.011) | 0.470 (0.002) | 2.448 (0.051) | 17.341 | 0.001 (0.001) |
| | INT-across | 1.094 (0.015) | 0.476 (0.011) | 0.473 (0.002) | 1.422 (0.036) | 10.073 | 0.107 (0.010) |
| | INT-across2 | -1.037 (0.024) | 0.761 (0.017) | 0.671 (0.003) | 1.654 (0.053) | 11.717 | 0.957 (0.006) |
| MI-regActive | INT-within | 3.258 (0.022) | 0.684 (0.015) | 0.582 (0.007) | 11.081 (0.128) | 78.497 | 0 (0) |
| | INT-across | 2.252 (0.050) | 1.587 (0.036) | 1.769 (0.048) | 7.587 (0.663) | 53.746 | 0.539 (0.016) |
| MI-redActive | INT-within | 3.433 (0.025) | 0.781 (0.017) | 0.605 (0.008) | 12.397 (0.181) | 87.819 | 0 (0) |
| | INT-across | 2.001 (0.026) | 0.818 (0.018) | 0.805 (0.009) | 4.674 (0.146) | 33.110 | 0.198 (0.013) |
| | INT-across2 | -0.950 (0.024) | 0.756 (0.017) | 0.654 (0.002) | 1.473 (0.049) | 10.435 | 1 (0) |
| *Auxiliary variable included in imputation model* | | | | | | | |
| MI-derPassive | INT-within | 0.319 (0.012) | 0.366 (0.008) | 0.378 (0.001) | 0.236 (0.009) | 1.672 | 1 (0) |
| | INT-across | 0.039 (0.012) | 0.379 (0.008) | 0.380 (0.001) | 0.145 (0.006) | 1.027 | 1 (0) |
| | INT-across2 | -1.104 (0.024) | 0.756 (0.017) | 0.677 (0.003) | 1.789 (0.057) | 12.673 | 0.954 (0.007) |
| MI-regActive | INT-within | 0.419 (0.012) | 0.394 (0.009) | 0.406 (0.001) | 0.331 (0.012) | 2.345 | 1 (0) |
| | INT-across | 1.642 (0.120) | 3.807 (0.085) | 3.516 (0.057) | 17.172 (1.684) | 121.645 | 0.991 (0.003) |
| MI-redActive | INT-within | 0.418 (0.012) | 0.381 (0.009) | 0.389 (0.001) | 0.320 (0.011) | 2.267 | 1 (0) |
| | INT-across | 0.045 (0.012) | 0.383 (0.009) | 0.388 (0.001) | 0.149 (0.007) | 1.056 | 1 (0) |

| | | | | | | |
|---|---|---|---|---|---|---|
| INT-across2 | -1.073 (0.023) | 0.741 (0.017) | 0.682 (0.002) | 1.700 (0.056) | 12.043 | 0.995 (0.002) |

Appendix

**Figure A1.** Illustration of bias in estimated treatment effects in 1,000 simulations *MI-regActive* when $X_2$ was missing under MNAR. The bias from each simulation when auxiliary variable was not included (black dots) was ranked from the lowest (left) to the highest (right). The red dots indicate the bias from the same simulation but when the auxiliary variable was included. We can see that, including an auxiliary variable reduces bias in most cases, except in the simulations with extremely high bias (right hand side of the plot). This explains why including an auxiliary variable increased the bootstrap variance as seen **Appendix Table A3**.

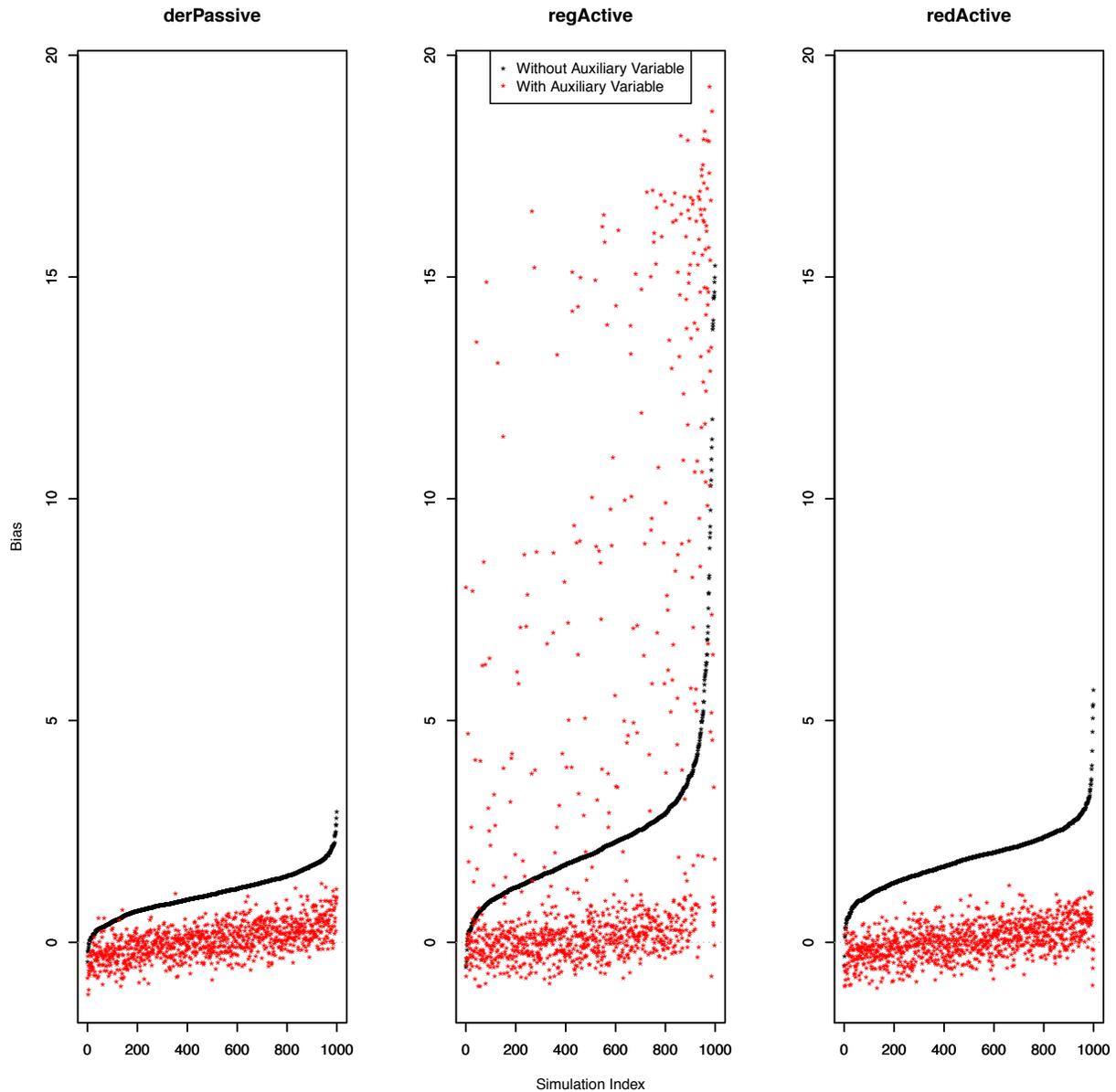

**Table A1.** Balance diagnosis and treatment effect estimation results using commonly applied missing data method before propensity score matching (PSM), in reference to applying PSM to full data without missingness (Monte Carlo standard errors in parentheses). CC = complete-case analysis, CVA = complete-variable analysis.

| MDM | Method | Bias | Stanford Error | | MSE | rMSE | Coverage | Percentage Matched | Post-Matching Standardized Difference | | |
| | | | Empirical | Robust | | | | | Underlying data | X1 | X2 |
|---|---|---|---|---|---|---|---|---|---|---|---|
| Full Data | NA | -0.006 (0.012) | 0.376 (0.008) | 0.380 (0.001) | 0.141 (0.006) | 1.000 | 0.950 (0.006) | 0.653 | - | 0.000 | 0.000 |
| MCAR | CC | 0.002 (0.011) | 0.513 (0.011) | 0.537 (0.001) | 0.262 (0.007) | 1.857 | 0.957 (0.006) | 0.653* | - | -0.001 | -0.004 |
|  | CVA | 5.058 (0.013) | 0.436 (0.01) | 0.442 (0) | 25.772 (0.132) | 182.565 | 0 (0) | 0.960 | - | 0 | 0.947 |
|  | Mean Imputation | 2.985 (0.014) | 0.461 (0.01) | 0.447 (0.001) | 9.122 (0.084) | 64.617 | 0 (0) | 0.803 | original data | 0 | 0.546 |
|  |  |  |  |  |  |  |  |  | imputed data | 0 | 0 |
|  | Missing Indicator | 2.973 (0.016) | 0.440 (0.01) | 0.446 (0.001) | 9.032 (0.098) | 63.981 | 0 (0) | 0.535 | full data | 0 | 0.546 |
|  |  |  |  |  |  |  |  |  | observed part | 0 | 0 |
|  |  |  |  |  |  |  |  |  | missing part | 0 | 0.946 |
| MAR1 | CC | -2.489 (0.017) | 0.821 (0.018) | 0.838 (0.002) | 6.869 (0.088) | 48.658 | 0.157 (0.012) | 1.000* | - | 0.002 | -0.001 |
|  | CVA | 5.059 (0.013) | 0.434 (0.01) | 0.443 (0) | 25.782 (0.131) | 182.634 | 0 (0) | 0.960 | - | 0 | 0.948 |
|  | Mean Imputation | 5.759 (0.017) | 0.573 (0.013) | 0.497 (0.004) | 33.498 (0.195) | 237.298 | 0 (0) | 0.735 | original data | 0 | 1.206 |
|  |  |  |  |  |  |  |  |  | imputed data | 0 | 0.095 |
|  | Missing Indicator | 5.201 (0.025) | 0.686 (0.015) | 0.674 (0.001) | 27.517 (0.267) | 194.929 | 0 (0) | 0.271 | full data | 0 | 0.936 |
|  |  |  |  |  |  |  |  |  | observed part | 0 | 0.034 |
|  |  |  |  |  |  |  |  |  | missing part | 0 | 1.517 |
| MAR2A | CC | -0.815 (0.015) | 0.726 (0.016) | 0.682 (0.002) | 1.191 (0.03) | 8.435 | 0.764 (0.013) | 1* | - | -0.001 | 0.001 |
|  | CVA | 5.059 (0.013) | 0.443 (0.01) | 0.442 (0) | 25.788 (0.134) | 182.680 | 0 (0) | 0.960 | - | 0 | 0.947 |
|  | Mean Imputation | 5.706 (0.017) | 0.574 (0.013) | 0.535 (0.002) | 32.886 (0.196) | 232.962 | 0 (0) | 0.602 | original data | 0 | 1.080 |
|  |  |  |  |  |  |  |  |  | imputed data | 0 | 0.044 |
|  | Missing Indicator | 5.528 (0.021) | 0.578 (0.013) | 0.568 (0.001) | 30.890 (0.237) | 218.819 | 0 (0) | 0.352 | full data | -0.001 | 0.954 |
|  |  |  |  |  |  |  |  |  | observed part | -0.001 | 0 |
|  |  |  |  |  |  |  |  |  | missing part | -0.001 | 1.704 |
| MNAR | CC | -1.084 (0.014) | 0.675 (0.015) | 0.682 (0.002) | 1.629 (0.033) | 11.538 | 0.631 (0.015) | 1* | - | -0.005 | -0.004 |
|  | CVA | 5.059 (0.013) | 0.435 (0.01) | 0.443 (0) | 25.783 (0.132) | 182.646 | 0 (0) | 0.960 | - | 0 | 0.947 |
|  | Mean Imputation | 5.689 (0.018) | 0.592 (0.013) | 0.559 (0.001) | 32.711 (0.202) | 231.723 | 0 (0) | 0.557 | original data | 0 | 1.078 |

| | | | | | | | | | |
|---|---|---|---|---|---|---|---|---|---|
| | | | | | | | imputed data | 0 | 0.045 |
| **Missing Indicator** | 5.534 (0.021) | 0.582 (0.013) | 0.577 (0.001) | 30.958 (0.239) | 219.305 | 0 (0) | 0.345 | full data | 0 | 0.962 |
| | | | | | | | | observed part | 0 | 0 |
| | | | | | | | | missing part | 0 | 1.725 |

*The percentage of treated patients being matched using CC is calculated from complete cases.

**Table A2.** Percentage of treated subjects matched and standardized difference of covariates after matching in various multiple imputation (MI) strategies. For *INT-across* and *INT-across2*, standardized differences were also calculated on the observed and missing part of the imputed data separately.

| | | | | Post-matching standardized difference | | | | | | | |
|---|---|---|---|---|---|---|---|---|---|---|---|
| | MI Strategy | | | Full Data | | Imputed Data | | Observed Part | | Missing Part | |
| MDM | Imputation | Integration | Percentage Matched | X1 | X2 | X1 | X2 | X1 | X2 | X1 | X2 |
| **MCAR** | | | | | | | | | | | |
| *Auxiliary variable not included in imputation model* | | | | | | | | | | | |
| | **MI-derPassive** | INT-within | 0.652 | 0 | 0.126 | 0 | 0 | - | - | - | - |
| | | INT-across | 0.633 | -0.002 | 0.008 | -0.002 | 0.006 | 0.036 | -0.114 | -0.046 | 0.125 |
| | | INT-across2 | 0.636 | -0.002 | 0.012 | -0.002 | 0.010 | 0.039 | -0.108 | -0.048 | 0.127 |
| | **MI-regActive** | INT-within | 0.791 | 0.128 | 0.600 | 0.128 | -0.127 | - | - | - | - |
| | | INT-across | 0.722 | -0.004 | 0.345 | -0.004 | 0.010 | 0.010 | 0.003 | -0.019 | 0.022 |
| | **MI-redActive** | INT-within | 0.792 | 0 | 0.656 | 0 | 0 | - | - | - | - |
| | | INT-across | 0.738 | 0 | 0.345 | 0 | 0.009 | 0.004 | -0.002 | -0.004 | 0.021 |
| | | INT-across2 | 0.782 | 0 | 0.430 | 0 | 0.045 | 0.001 | 0.001 | -0.002 | 0.079 |
| *Auxiliary variable included in imputation model* | | | | | | | | | | | |
| | **MI-derPassive** | INT-within | 0.655 | 0 | 0.024 | 0 | 0 | - | - | - | - |
| | | INT-across | 0.652 | 0.006 | -0.008 | 0.006 | -0.001 | -0.115 | -0.088 | 0.098 | 0.066 |
| | | INT-across2 | 0.653 | 0.010 | -0.011 | 0.010 | -0.003 | -0.123 | -0.087 | 0.110 | 0.062 |
| | **MI-regActive** | INT-within | 0.659 | 0.006 | 0.040 | 0.006 | -0.011 | - | - | - | - |
| | | INT-across | 0.653 | -0.028 | 0.025 | -0.028 | 0.010 | -0.122 | 0.053 | 0.049 | -0.039 |
| | **MI-redActive** | INT-within | 0.659 | 0 | 0.051 | 0 | 0 | - | - | - | - |
| | | INT-across | 0.653 | -0.031 | 0.029 | -0.031 | 0.014 | -0.203 | 0.086 | 0.110 | -0.059 |
| | | INT-across2 | 0.652 | -0.043 | 0.040 | -0.043 | 0.026 | -0.214 | 0.112 | 0.100 | -0.061 |
| **MAR1** | | | | | | | | | | | |
| *Auxiliary variable not included in imputation model* | | | | | | | | | | | |
| | **MI-derPassive** | INT-within | 0.656 | 0 | 0.114 | 0 | 0 | - | - | - | - |
| | | INT-across | 0.617 | 0.002 | -0.028 | 0.002 | -0.015 | -0.748 | -1.105 | 0.205 | 0.647 |
| | | INT-across2 | 0.621 | 0.008 | -0.034 | 0.008 | -0.021 | -0.746 | -1.109 | 0.201 | 0.645 |
| | **MI-regActive** | INT-within | 0.877 | 0.368 | 0.752 | 0.368 | 0.322 | NA | NA | NA | NA |
| | | INT-across | 0.666 | 0.562 | 0.531 | 0.562 | 0.344 | -0.160 | -0.397 | 0.714 | 0.708 |
| | **MI-redActive** | INT-within | 0.772 | 0 | 0.483 | 0 | -0.006 | - | - | - | - |
| | | INT-across | 0.628 | 0.002 | 0.031 | 0.002 | -0.178 | -0.746 | -1.013 | 0.241 | 0.466 |
| | | INT-across2 | 0.670 | 0.003 | 0.219 | 0.003 | -0.029 | -0.780 | -0.84 | 0.260 | 0.517 |
| *Auxiliary variable included in imputation model* | | | | | | | | | | | |
| | **MI-derPassive** | INT-within | 0.666 | 0 | 0.065 | 0 | -0.001 | - | - | - | - |
| | | INT-across | 0.647 | 0.026 | -0.031 | 0.026 | -0.069 | -0.895 | -1.010 | 0.467 | 0.404 |
| | | INT-across2 | 0.652 | 0.022 | -0.027 | 0.022 | -0.060 | -0.953 | -0.974 | 0.519 | 0.375 |

|  |  |  |  |  |  |  |  |  |  |  |
|---|---|---|---|---|---|---|---|---|---|---|
| | **MI-regActive** | **INT-within** | 0.785 | 0.351 | 0.420 | 0.351 | 0.338 | - | - | - | - |
| | | **INT-across** | 0.745 | 0.386 | 0.382 | 0.386 | 0.321 | -0.672 | -0.663 | 0.856 | 0.741 |
| | **MI-redActive** | **INT-within** | 0.669 | 0 | 0.085 | 0 | -0.001 | - | - | - | - |
| | | **INT-across** | 0.643 | 0.012 | -0.015 | 0.012 | -0.069 | -0.921 | -1.008 | 0.460 | 0.405 |
| | | **INT-across2** | 0.651 | 0.013 | -0.016 | 0.013 | -0.061 | -0.972 | -0.982 | 0.506 | 0.384 |
| **MAR2A** | | | | | | | | | | | |
| *Auxiliary variable not included in imputation model* | | | | | | | | | | | |
| | **MI-derPassive** | **INT-within** | 0.754 | 0 | 0.379 | 0 | 0 | - | - | - | - |
| | | **INT-across** | 0.686 | 0 | 0.146 | 0 | -0.017 | -0.390 | -1.182 | 0.226 | 0.558 |
| | | **INT-across2** | 0.695 | 0 | 0.171 | 0 | -0.016 | -0.392 | -1.185 | 0.212 | 0.563 |
| | **MI-regActive** | **INT-within** | 0.881 | 0.076 | 0.698 | 0.076 | 0.078 | - | - | - | - |
| | | **INT-across** | 0.571 | 0.026 | 0.428 | 0.026 | 0.021 | -0.038 | -0.472 | 0.020 | 0.204 |
| | **MI-redActive** | **INT-within** | 0.863 | 0 | 0.656 | 0 | 0 | - | - | - | - |
| | | **INT-across** | 0.734 | 0 | 0.348 | 0 | -0.095 | -0.331 | -0.990 | 0.107 | 0.221 |
| | | **INT-across2** | 0.780 | 0 | 0.396 | 0 | -0.090 | -0.403 | -1.061 | 0.140 | 0.245 |
| *Auxiliary variable included in imputation model* | | | | | | | | | | | |
| | **MI-derPassive** | **INT-within** | 0.669 | 0 | 0.071 | 0 | -0.001 | - | - | - | - |
| | | **INT-across** | 0.652 | -0.014 | 0.013 | -0.014 | -0.030 | -0.441 | -1.283 | 0.294 | 0.546 |
| | | **INT-across2** | 0.653 | -0.025 | 0.023 | -0.025 | -0.019 | -0.473 | -1.257 | 0.308 | 0.541 |
| | **MI-regActive** | **INT-within** | 0.729 | 0.122 | 0.248 | 0.122 | 0.166 | - | - | - | - |
| | | **INT-across** | 0.701 | 0.131 | 0.177 | 0.131 | 0.126 | -0.374 | -1.061 | 0.510 | 0.614 |
| | **MI-redActive** | **INT-within** | 0.672 | 0 | 0.084 | 0 | -0.001 | - | - | - | - |
| | | **INT-across** | 0.652 | -0.018 | 0.017 | -0.018 | -0.032 | -0.476 | -1.230 | 0.341 | 0.480 |
| | | **INT-across2** | 0.652 | -0.026 | 0.024 | -0.026 | -0.023 | -0.506 | -1.203 | 0.361 | 0.465 |
| **MNAR** | | | | | | | | | | | |
| *Auxiliary variable not included in imputation model* | | | | | | | | | | | |
| | **MI-derPassive** | **INT-within** | 0.754 | 0 | 0.377 | 0 | 0 | - | - | - | - |
| | | **INT-across** | 0.686 | 0 | 0.143 | 0 | -0.017 | -0.413 | -1.245 | 0.233 | 0.571 |
| | | **INT-across2** | 0.694 | 0 | 0.168 | 0 | -0.017 | -0.416 | -1.247 | 0.219 | 0.576 |
| | **MI-regActive** | **INT-within** | 0.888 | 0.117 | 0.733 | 0.117 | 0.107 | - | - | NA | NA |
| | | **INT-across** | 0.524 | 0.041 | 0.437 | 0.041 | 0.032 | 0.066 | -0.410 | -0.045 | 0.179 |
| | **MI-redActive** | **INT-within** | 0.863 | 0 | 0.652 | 0 | 0 | - | - | - | - |
| | | **INT-across** | 0.729 | 0 | 0.329 | 0 | -0.107 | -0.346 | -1.080 | 0.093 | 0.229 |
| | | **INT-across2** | 0.776 | 0 | 0.381 | 0 | -0.099 | -0.418 | -1.142 | 0.132 | 0.256 |
| *Auxiliary variable included in imputation model* | | | | | | | | | | | |
| | **MI-derPassive** | **INT-within** | 0.671 | 0 | 0.079 | 0 | -0.001 | - | - | - | - |
| | | **INT-across** | 0.652 | -0.012 | 0.011 | -0.012 | -0.039 | -0.457 | -1.339 | 0.301 | 0.542 |
| | | **INT-across2** | 0.652 | -0.022 | 0.021 | -0.022 | -0.029 | -0.488 | -1.310 | 0.317 | 0.531 |
| | **MI-regActive** | **INT-within** | 0.747 | 0.204 | 0.300 | 0.204 | 0.208 | - | - | - | - |
| | | **INT-across** | 0.709 | 0.240 | 0.246 | 0.240 | 0.184 | -0.294 | -1.027 | 0.623 | 0.636 |

| | | | | | | | | | | |
|---|---|---|---|---|---|---|---|---|---|---|
| **MI-redActive** | **INT-within** | 0.675 | 0 | 0.096 | 0 | -0.001 | NA | NA | NA | NA |
| | **INT-across** | 0.652 | -0.012 | 0.011 | -0.012 | -0.047 | -0.494 | -1.281 | 0.364 | 0.455 |
| | **INT-across2** | 0.652 | -0.020 | 0.019 | -0.020 | -0.037 | -0.528 | -1.251 | 0.387 | 0.441 |

**Table A3.** Bias, standard error mean squared error (MSE), relative mean squared error (rMSE), and coverage (calculated using bootstrap standard error) results using various multiple imputation (MI) strategies in MCAR, MAR1, and MNAR. (Monte Carlo standard errors in parentheses).

| MI Strategy | | | Standard Error | | | | |
| --- | --- | --- | --- | --- | --- | --- | --- |
| Imputation | Integration | Bias | Empirical | Bootstrap | MSE | rMSE | Coverage |
| **MCAR** | | | | | | | |
| *Auxiliary variable not included in imputation model* | | | | | | | |
| MI-derPassive | INT-within | -0.024 (0.012) | 0.381 (0.009) | 0.395 (0.001) | 0.146 (0.006) | 1.034 | 1 (0) |
| | INT-across | -0.796 (0.013) | 0.424 (0.009) | 0.418 (0.001) | 0.814 (0.022) | 5.766 | 0.569 (0.016) |
| | INT-across2 | 0.008 (0.017) | 0.534 (0.012) | 0.539 (0.002) | 0.285 (0.013) | 2.019 | 1 (0) |
| MI-regActive | INT-within | 3.224 (0.015) | 0.489 (0.011) | 0.508 (0.002) | 10.636 (0.103) | 75.345 | 0 (0) |
| | INT-across | 0.870 (0.028) | 0.882 (0.020) | 0.813 (0.004) | 1.534 (0.055) | 10.867 | 1 (0) |
| MI-redActive | INT-within | 3.309 (0.013) | 0.405 (0.009) | 0.438 (0.001) | 11.113 (0.087) | 78.724 | 0 (0) |
| | INT-across | 0.851 (0.029) | 0.926 (0.021) | 0.807 (0.004) | 1.580 (0.057) | 11.193 | 1 (0) |
| | INT-across2 | 0.012 (0.017) | 0.541 (0.012) | 0.536 (0.001) | 0.293 (0.013) | 2.076 | 1 (0) |
| *Auxiliary variable included in imputation model* | | | | | | | |
| MI-derPassive | INT-within | 0.095 (0.011) | 0.359 (0.008) | 0.359 (0.001) | 0.138 (0.006) | 0.978 | 1 (0) |
| | INT-across | 0.046 (0.012) | 0.388 (0.009) | 0.390 (0.001) | 0.152 (0.007) | 1.077 | 1 (0) |
| | INT-across2 | -0.005 (0.017) | 0.55 (0.012) | 0.547 (0.001) | 0.303 (0.013) | 2.146 | 1 (0) |
| MI-regActive | INT-within | 0.241 (0.011) | 0.358 (0.008) | 0.363 (0.001) | 0.186 (0.008) | 1.318 | 1 (0) |
| | INT-across | 0.231 (0.012) | 0.384 (0.009) | 0.405 (0.001) | 0.200 (0.009) | 1.417 | 1 (0) |
| MI-redActive | INT-within | 0.247 (0.011) | 0.356 (0.008) | 0.361 (0.001) | 0.188 (0.008) | 1.332 | 1 (0) |
| | INT-across | 0.221 (0.012) | 0.387 (0.009) | 0.398 (0.001) | 0.198 (0.008) | 1.403 | 1 (0) |
| | INT-across2 | 0.003 (0.017) | 0.534 (0.012) | 0.551 (0.001) | 0.285 (0.013) | 2.019 | 1 (0) |
| **MAR1** | | | | | | | |
| *Auxiliary variable not included in imputation model* | | | | | | | |
| MI-derPassive | INT-within | 0.038 (0.014) | 0.454 (0.01) | 0.446 (0.001) | 0.207 (0.011) | 1.466 | 1 (0) |
| | INT-across | -0.176 (0.016) | 0.497 (0.011) | 0.472 (0.002) | 0.278 (0.012) | 1.969 | 1 (0) |
| | INT-across2 | -2.859 (0.024) | 0.770 (0.017) | 0.757 (0.003) | 8.764 (0.14) | 62.083 | 0 (0) |
| MI-regActive | INT-within | 2.471 (0.043) | 1.345 (0.03) | 1.463 (0.007) | 7.913 (0.174) | 56.055 | 0.823 (0.012) |
| | INT-across | 5.275 (0.173) | 5.473 (0.122) | 4.719 (0.033) | 57.746 (2.605) | 409.068 | 0.915 (0.009) |
| MI-redActive | INT-within | 2.557 (0.045) | 1.414 (0.032) | 1.405 (0.006) | 8.534 (0.183) | 60.454 | 0.719 (0.014) |
| | INT-across | 0.465 (0.032) | 1.013 (0.023) | 0.926 (0.005) | 1.240 (0.076) | 8.784 | 1 (0) |
| | INT-across2 | -2.633 (0.025) | 0.792 (0.018) | 0.772 (0.003) | 7.558 (0.132) | 53.540 | 0 (0) |
| *Auxiliary variable included in imputation model* | | | | | | | |
| MI-derPassive | INT-within | 0.319 (0.012) | 0.395 (0.009) | 0.376 (0.001) | 0.258 (0.01) | 1.828 | 1 (0) |
| | INT-across | 0.067 (0.013) | 0.398 (0.009) | 0.385 (0.001) | 0.163 (0.008) | 1.155 | 1 (0) |
| | INT-across2 | -2.904 (0.024) | 0.771 (0.017) | 0.744 (0.003) | 9.028 (0.141) | 63.954 | 0 (0) |
| MI-regActive | INT-within | 0.502 (0.014) | 0.449 (0.01) | 0.465 (0.001) | 0.453 (0.022) | 3.209 | 1 (0) |
| | INT-across | 4.096 (0.176) | 5.572 (0.125) | 4.702 (0.040) | 47.793 (2.478) | 338.561 | 0.948 (0.007) |
| MI-redActive | INT-within | 0.461 (0.013) | 0.409 (0.009) | 0.421 (0.001) | 0.380 (0.016) | 2.692 | 1 (0) |
| | INT-across | 0.059 (0.013) | 0.418 (0.009) | 0.406 (0.001) | 0.178 (0.008) | 1.261 | 1 (0) |
| | INT-across2 | -2.860 (0.024) | 0.770 (0.017) | 0.768 (0.003) | 8.773 (0.14) | 62.147 | 0 (0) |

**MNAR**

*Auxiliary variable not included in imputation model*

| | | | | | | | |
|---|---|---|---|---|---|---|---|
| MI-derPassive | INT-within | 1.487 (0.016) | 0.515 (0.012) | 0.470 (0.002) | 2.476 (0.052) | 17.540 | 0.001 (0.001) |
| | INT-across | 1.096 (0.015) | 0.485 (0.011) | 0.473 (0.002) | 1.437 (0.037) | 10.180 | 0.103 (0.01) |
| | INT-across2 | -1.345 (0.021) | 0.675 (0.015) | 0.658 (0.003) | 2.264 (0.06) | 16.038 | 0.332 (0.015) |
| MI-regActive | INT-within | 3.384 (0.016) | 0.498 (0.011) | 0.541 (0.004) | 11.701 (0.111) | 82.889 | 0 (0) |
| | INT-across | 2.371 (0.062) | 1.953 (0.044) | 2.255 (0.050) | 9.430 (0.793) | 66.801 | 0.666 (0.015) |
| MI-redActive | INT-within | 3.404 (0.016) | 0.519 (0.012) | 0.539 (0.007) | 11.860 (0.119) | 84.015 | 0 (0) |
| | INT-across | 1.878 (0.021) | 0.661 (0.015) | 0.719 (0.007) | 3.963 (0.092) | 28.074 | 0.1 (0.009) |
| | INT-across2 | -1.261 (0.02) | 0.637 (0.014) | 0.646 (0.002) | 1.996 (0.054) | 14.139 | 0.483 (0.016) |

*Auxiliary variable included in imputation model*

| | | | | | | | |
|---|---|---|---|---|---|---|---|
| MI-derPassive | INT-within | 0.408 (0.012) | 0.377 (0.008) | 0.380 (0.001) | 0.309 (0.011) | 2.189 | 1 (0) |
| | INT-across | 0.042 (0.012) | 0.382 (0.009) | 0.378 (0.001) | 0.148 (0.007) | 1.048 | 1 (0) |
| | INT-across2 | -1.367 (0.021) | 0.679 (0.015) | 0.667 (0.003) | 2.33 (0.062) | 16.506 | 0.323 (0.015) |
| MI-regActive | INT-within | 0.490 (0.013) | 0.398 (0.009) | 0.406 (0.001) | 0.398 (0.015) | 2.819 | 1 (0) |
| | INT-across | 2.505 (0.159) | 5.024 (0.112) | 4.295 (0.054) | 31.492 (2.419) | 223.087 | 0.957 (0.006) |
| MI-redActive | INT-within | 0.480 (0.012) | 0.387 (0.009) | 0.392 (0.001) | 0.380 (0.014) | 2.692 | 1 (0) |
| | INT-across | 0.045 (0.013) | 0.396 (0.009) | 0.389 (0.001) | 0.159 (0.007) | 1.126 | 1 (0) |
| | INT-across2 | -1.325 (0.021) | 0.661 (0.015) | 0.673 (0.002) | 2.191 (0.058) | 15.521 | 0.455 (0.016) |

**Table A4.** Bias, standard error mean squared error (MSE), relative mean squared error (rMSE), and coverage calculated using *MI-regPassive* and *MI-derPassive* when $X_2$ was missing MAR2B and $Z_{ps}$ was missing under various missing data mechanisms (aux_MCAR, aux_MAR1, and aux_MAR2). Monte Carlo standard errors are in parentheses.

| MI Strategy | | | Standard Error | | | | |
| --- | --- | --- | --- | --- | --- | --- | --- |
| Imputation | Integration | Bias | Empirical | Bootstrap | MSE | rMSE | Coverage |
| **Fully observed aux** | | | | | | | |
| MI-derPassive | INT-within | -0.019 (0.012) | 0.381 (0.009) | 0.378 (0.001) | 0.146 (0.006) | 1.032 | 1 (0) |
| | INT-across | -0.205 (0.012) | 0.394 (0.009) | 0.381 (0.001) | 0.197 (0.008) | 1.394 | 1 (0) |
| | INT-across2 | -1.016 (0.021) | 0.652 (0.015) | 0.682 (0.003) | 1.457 (0.048) | 10.319 | 0.993 (0.003) |
| **aux_MCAR** | | | | | | | |
| *Default imputation order* | | | | | | | |
| MI-derPassive | INT-within | 0.085 (0.012) | 0.381 (0.009) | 0.550 (0.006) | 0.152 (0.007) | 1.079 | 1 (0) |
| | INT-across | -0.152 (0.013) | 0.402 (0.009) | 0.499 (0.004) | 0.184 (0.008) | 1.304 | 1 (0) |
| | INT-across2 | -0.982 (0.021) | 0.671 (0.015) | 0.683 (0.002) | 1.415 (0.048) | 10.024 | 1 (0) |
| MI-regPassive | INT-within | 0.087 (0.012) | 0.378 (0.008) | 0.533 (0.006) | 0.150 (0.007) | 1.065 | 1 (0) |
| | INT-across | -0.143 (0.013) | 0.413 (0.009) | 0.489 (0.003) | 0.191 (0.008) | 1.352 | 1 (0) |
| | INT-across2 | -0.996 (0.021) | 0.666 (0.015) | 0.685 (0.002) | 1.436 (0.046) | 10.173 | 1 (0) |
| *Reverse imputation order* | | | | | | | |
| MI-derPassive | INT-within | 0.028 (0.012) | 0.382 (0.009) | 0.550 (0.006) | 0.147 (0.006) | 1.040 | 1 (0) |
| | INT-across | -0.191 (0.012) | 0.391 (0.009) | 0.500 (0.004) | 0.189 (0.008) | 1.340 | 1 (0) |
| | INT-across2 | -1.01 (0.021) | 0.671 (0.015) | 0.682 (0.002) | 1.471 (0.047) | 10.418 | 0.999 (0.001) |
| MI-regPassive | INT-within | -0.196 (0.012) | 0.372 (0.008) | 0.532 (0.006) | 0.177 (0.008) | 1.251 | 1 (0) |
| | INT-across | -0.232 (0.013) | 0.405 (0.009) | 0.489 (0.003) | 0.218 (0.009) | 1.543 | 1 (0) |
| | INT-across2 | -1.002 (0.021) | 0.668 (0.015) | 0.684 (0.002) | 1.450 (0.045) | 10.273 | 1 (0) |
| **aux_MAR1** | | | | | | | |
| *Default imputation order* | | | | | | | |
| MI-derPassive | INT-within | 0.185 (0.012) | 0.393 (0.009) | 0.550 (0.006) | 0.189 (0.008) | 1.337 | 1 (0) |
| | INT-across | 0.011 (0.013) | 0.414 (0.009) | 0.499 (0.004) | 0.171 (0.008) | 1.211 | 1 (0) |
| | INT-across2 | -1.017 (0.021) | 0.651 (0.015) | 0.684 (0.002) | 1.458 (0.045) | 10.329 | 1 (0) |
| MI-regPassive | INT-within | 0.205 (0.013) | 0.396 (0.009) | 0.533 (0.006) | 0.199 (0.008) | 1.410 | 1 (0) |
| | INT-across | 0.014 (0.013) | 0.414 (0.009) | 0.488 (0.003) | 0.171 (0.008) | 1.213 | 1 (0) |
| | INT-across2 | -1.007 (0.021) | 0.656 (0.015) | 0.685 (0.002) | 1.444 (0.045) | 10.231 | 1 (0) |
| *Reverse imputation order* | | | | | | | |
| MI-derPassive | INT-within | 0.091 (0.012) | 0.391 (0.009) | 0.550 (0.006) | 0.161 (0.007) | 1.140 | 1 (0) |
| | INT-across | -0.069 (0.013) | 0.408 (0.009) | 0.500 (0.004) | 0.171 (0.007) | 1.213 | 1 (0) |
| | INT-across2 | -0.993 (0.021) | 0.661 (0.015) | 0.684 (0.002) | 1.422 (0.045) | 10.071 | 1 (0) |
| MI-regPassive | INT-within | -0.129 (0.012) | 0.379 (0.008) | 0.534 (0.006) | 0.160 (0.007) | 1.135 | 1 (0) |
| | INT-across | -0.087 (0.013) | 0.403 (0.009) | 0.489 (0.003) | 0.170 (0.008) | 1.201 | 1 (0) |
| | INT-across2 | -0.982 (0.021) | 0.650 (0.015) | 0.685 (0.002) | 1.385 (0.043) | 9.813 | 1 (0) |

**aux_MAR2**

*Default imputation order*

| | | | | | | | |
|---|---|---|---|---|---|---|---|
| MI-derPassive | INT-within | -0.060 (0.012) | 0.390 (0.009) | 0.550 (0.006) | 0.155 (0.006) | 1.100 | 1 (0) |
| | INT-across | -0.374 (0.013) | 0.412 (0.009) | 0.500 (0.004) | 0.310 (0.012) | 2.194 | 1 (0) |
| | INT-across2 | -0.994 (0.021) | 0.660 (0.015) | 0.683 (0.002) | 1.422 (0.044) | 10.076 | 1 (0) |
| MI-regPassive | INT-within | -0.049 (0.013) | 0.396 (0.009) | 0.533 (0.006) | 0.159 (0.006) | 1.128 | 1 (0) |
| | INT-across | -0.369 (0.013) | 0.408 (0.009) | 0.489 (0.003) | 0.302 (0.012) | 2.141 | 1 (0) |
| | INT-across2 | -1.001 (0.021) | 0.656 (0.015) | 0.685 (0.002) | 1.431 (0.043) | 10.137 | 0.999 (0.001) |

*Reverse imputation order*

| | | | | | | | |
|---|---|---|---|---|---|---|---|
| MI-derPassive | INT-within | -0.068 (0.012) | 0.390 (0.009) | 0.549 (0.006) | 0.157 (0.006) | 1.111 | 1 (0) |
| | INT-across | -0.369 (0.013) | 0.412 (0.009) | 0.500 (0.004) | 0.305 (0.012) | 2.163 | 1 (0) |
| | INT-across2 | -0.996 (0.021) | 0.652 (0.015) | 0.683 (0.002) | 1.416 (0.043) | 10.032 | 0.999 (0.001) |
| MI-regPassive | INT-within | -0.308 (0.012) | 0.391 (0.009) | 0.533 (0.006) | 0.248 (0.010) | 1.754 | 1 (0) |
| | INT-across | -0.414 (0.013) | 0.411 (0.009) | 0.489 (0.003) | 0.340 (0.013) | 2.410 | 1 (0) |
| | INT-across2 | -0.987 (0.021) | 0.665 (0.015) | 0.685 (0.002) | 1.417 (0.043) | 10.039 | 1 (0) |